\documentclass{fizik} 
\vol{}
\pyear{}
\received{}

\newcommand{\ben}{\begin{enumerate}}
\newcommand{\een}{\end{enumerate}}
\newcommand{\be}{\begin{equation}}
\newcommand{\ee}{\end{equation}}
\newcommand{\bse}{\begin{subequation}}
\newcommand{\ese}{\end{subequation}}
\newcommand{\bea}{\begin{eqnarray}}
\newcommand{\eea}{\end{eqnarray}}
\newcommand{\bc}{\begin{center}}
\newcommand{\ec}{\end{center}}

\author[OMOTE AND KAMEFUCHI]
        {
        {\bf  Minoru Omote }\\
        {\it  Department of Physics , Keio University , }\\
        {\it  Hiyoshi , Yokohama  }\\
        {\bf  Susumu Kamefuchi }\\
        {\it  Atomic Energy Research Institute , Nihon University ,  }\\
        {\it  Kanda-Surugadai , Tokyo  }\\

        }
\title{ \hspace{0.8cm} Inertial Force, Equivalence Principle and  Quantum Mechanics }

\begin{document}
\maketitle

\begin{abstract}
On the basis of a manifestly covariant formalism of non-relativistic quantum mechanics in general coordinate systems, proposed by us recently, we derive  general expressions for inertial forces. The results enable us further to discuss, and to explain the validity of, the equivalence principle in non-relativistic quantum mechanics.
 
\end{abstract}

\section{Introduction}
In the usual formulation of non-relativistic quantum mechanics (NRQM) basic quantities such as fields or probability amplitudes are taken to be projective representations of the 4-dimensional Galilei group $G_{4}$, and the manifest covariance of the formalism is thereby impaired. As shown previously [1], this shortcoming can be removed, however, by employing, instead of $G_{5}$, a 5-dimensional form $G_{5}$ for Galilei transformations [2]. Here, basic quantities can be given as vector representations of the group $G_{5}$, so that the manifest covariance is achieved, just as Lorentz covariance is in relativistic quantum mechanics (RQM).

In a recent paper [3] the result of [1] has been extended by considering more general transformations than $G_{5}$, i.e., those which combine inertial with non-inertial coordinate systems. In this way we have obtained generally covariant equations for basic quantities that should hold in arbitrary coordinate systems, inertial or non-inertial.

Naturally, the equations thus obtained contain the terms corresponding to inertial forces. And it is the purpose of this paper to discuss such terms, placing emphasis upon the problem of Einstein's equivalence principle. Our conclusion then is that NRQM is compatible with this principle. 

In what follows we shall exclusively be concerned with field equations for a non-relativistic field $\psi(\vec{x},t)$. However, when $\psi$ satisfies a basic equation of the Schr$\ddot{{\rm o}}$dinger type such as (9) below, the probability amplitude $\varphi(\vec{x},t)$ for a single particle, arising from the quantized $\psi$, does also satisfy one and the same equation. Thus, all arguments about covariance of $\psi$ apply as well to $\varphi$.

\section{5-dimensional Transformations: $G_{5}$ and $G_{5}^{\prime}$}
We begin by noticing that the Lagrangian ${\cal L}= (1/2)m \dot{\vec{x}}^{2}$ of a free particle of mass m is not   invariant under $G_{4}$:  $\vec{x} \to \vec{x}^{\prime}=R\vec{x}-\vec{v}t, t^{\prime}=t$ with $R$ being a $3\times 3$ orthogonal matrix. However, $\bar{{\cal L}}\equiv {\cal L}-m \dot{s}$ remains invariant, provided the new variable $s$ transforms under $G_{4}$ as $s \to  s^{\prime}=s+f$ with $f(\vec{x},t) \equiv -\vec{v} \cdot (R\vec{x}) +(1/2)\vec{v}^{2}t.$ Thus, instead of $G_{4}$ we employ $G_{5}$, a central extension of $G_{4}$, such that 
\begin{equation}
\begin{array}{l}
x^{\prime i}=R^{i}~_{j}~x^{j}-\frac{v^{i}}{u}x^{4},\hspace{1cm} x^{\prime 4}=x^{4},  \\
x^{\prime 5}=x^{5}-\frac{v_{i}}{u}(R^{i}~_{j} x^{j})+\frac{1}{2}\frac{\vec{v}^{2}}{u^{2}}x^{4} ,
\end{array}
\end{equation}
where $(x^{1},x^{2},x^{3}) \equiv \vec{x},~x^{4} \equiv ut, x^{5} \equiv s/u$ with $[u]=[v]$, or $x^{\prime \mu}\equiv \Lambda^{\mu}~_{\nu} x^{\nu}$. Under (1) $\eta_{\mu \nu}x^{\mu}x^{\nu}=\vec{x}^{2}-2ts$ is invariant, where the metric tensor $\parallel \eta_{\mu \nu}\parallel=\parallel \eta^{\mu \nu}\parallel$ is  such that $\eta_{ij}=\delta_{ij} (i,j=1,2,3), \eta_{45}=\eta_{54}=-1$ and others $=0$. Note that $\eta_{\mu\nu}x^{\mu}x^{\nu}$ with $x_{\pm}\equiv (x^{4} \pm x^{5})/\sqrt{2}$ leads to $\vec{x}^{2}+\vec{x}_{-}^{2}-x_{+}^{2}$, and $\eta_{\mu\nu}p^{\mu}p^{\nu}=0$ with $p^{\mu} \equiv (\vec{p},mu,E/u)$ to $E=\vec{p}^{2}/2m$.

Thus, if the original coordinate system S$_{0}$ with coordinates $x^{\mu}$ is an inertial system, so is the transformed system S with coordinates $x^{\prime \mu}$. Our basic assumption then is that NRQM in inertial systems be invariant under $G_{5}$.

We next generalize $G_{4}$ to $G_{4}^{\prime}$ such as $\vec{x}^{\prime}=R(t)\vec{x}+\vec{A}(t),~t^{\prime}=t$, where $R$ and a vector $\vec{A}$ are taken to be $t$-dependent. The corresponding transformation rule of $s$ can again be determined  by equating the total-time-derivative term in  ( ${\cal L}(\vec{x},t)=){\cal L}[\vec{x}(\vec{x}^{\prime},t^{\prime}),t^{\prime}]$ to $-m(\dot{s}^{\prime}-\dot{s})$. Thus, instead of (1) we now consider $G_{5}^{\prime}$ such that 
\begin{equation}
\begin{array}{l}
x^{\prime i}=R^{i}~_{j}x^{j}+A^{i},   \hspace{1cm} x^{\prime 4}=x^{4},  \\
x^{\prime 5}=x^{5}+\frac{1}{u}\dot{\tilde{A}}_{j}x^{j}+\frac{1}{u}\tilde{A}_{j}\dot{\tilde{A}}~^{j}-\frac{1}{2u^{2}}\int_{0}^{x^{4}}\dot{\tilde{A}}_{j}(\tau )\dot{\tilde{A}}~^{j}(\tau)d \tau,   
\end{array} 
\end{equation}
where $\tilde{A}_{i} \equiv R^{j}~_{i}A_{j}$. Obviously, this  rule for $x^{5} \to x^{\prime 5}$  is not unique , and the above is the simplest choice: the  arbitrariness arising from this  will not affect the final results, however. The coordinate system S obtained from S$_{0}$ by (2) is non-inertial in general. Thus non-inertial systems will be restricted hereafter to those specified by $R(t)$ and $\vec{A}(t)$.

In the coordinate system S the metric tensor $g^{\prime \mu\nu}$ is given by
\begin{equation}
\begin{array}{lll}
g^{\prime ij}=\delta^{ij}, & g^{\prime i4}=0, & g^{\prime i5}=-\frac{1}{u}\dot{R}^{i}~_{j}R_{k}~^{j}x^{\prime k},  \\
g^{\prime 44}=0, & g^{\prime 45}=-1, & g^{\prime 55}=-\frac{2}{u^{2}}R_{ji}~\ddot{\tilde{A}}~^{i}~x^{\prime j};
\end{array}
\end{equation}
and the affine connection $\Gamma^{\prime \lambda}_{\mu \nu}(x^{\prime})$ by
\begin{equation}
\begin{array}{ll}
\Gamma^{\prime i}_{4j}=\frac{1}{u}R^{i}~_{k}\dot{ R}_{j}~^{k},& \Gamma^{\prime i}_{44}=\frac{1}{u^{2}}R^{i}~_{j}\ddot{ R}_{k}~^{j}x^{\prime k}-\frac{1}{u^{2}}R^{i}~_{j}\ddot{\tilde{A}} ~^{j},  \\
\Gamma^{\prime 5}_{4i}=-\frac{1}{u^{2}}R_{ij}\ddot{\tilde{A}}~^{j}, & \Gamma^{\prime 5}_{44 }=-\frac{2}{u^{3}}\dot{R}_{ij}~\ddot{\tilde{A}}~^{j}~x^{\prime i}-\frac{1}{u^{3}}~R_{ij}\stackrel{\cdots}{\tilde{A}}~^{j}~x^{\prime i},   
\end{array}
\end{equation}
\[{\rm others}=0 . \]

\section{Generally Covariant Field Equations}
For illustration we consider in the following a scalar field $\phi(x)$ and a spinor field $\chi(x)$: the generalization to the case of Bargmann-Wigner fields of higher spins is straightforward.

As shown in [3], a Klein-Gordon type equation $\eta^{\mu\nu}\partial_{\mu}\partial_{\nu}\phi(x)=0$ for $\phi(x)$ in S$_{0}$ takes, in S, the following form:
\begin{equation}
g^{\prime \mu\nu}\partial^{\prime}_{\mu}\partial^{\prime}_{\nu}\phi^{\prime}(x^{\prime})=0 .
\end{equation}
Further, this should be supplemented by a subsidiary condition of invariant form:
\[\hspace{4.5cm} 
(i\hbar\partial^{\prime}_{5}-m u)\phi^{\prime}(x^{\prime})=0 , \hspace{5cm} (5^{\prime}) 
\]
with which to eliminate the extra degree of freedom.

On the other hand, a spinor field $\chi(x)$ is to behave as a scalar under (2), i.e., $\chi^{\prime}(x^{\prime})=\chi(x)$, and a Dirac-type equation $\gamma^{\mu}\partial_{\mu}\chi(x)=0$ in S$_{0}$ is generalized, in S, to
\begin{equation}
\gamma^{\prime \mu}(x^{\prime})\bigl(\partial^{\prime}_{\mu}+\Gamma^{\prime}_{\mu}(x^{\prime})\bigr)\chi^{\prime}(x^{\prime})=0.
\end{equation}
The subsidiary condition for $\chi^{\prime}(x^{\prime})$ can be of the same form as (5$^{\prime}$).

In the above the $\gamma^{\prime}$-matrices and the spin connection $\Gamma^{\prime}_{\mu}(x^{\prime})$ are given, in terms of the f$\ddot{{\rm u}}$nfbein $h^{\mu}_{a}(x^{\prime})$ and of $4\times 4$ $\gamma$-matrices satisfying $\gamma^{a}\gamma^{b}+\gamma^{b}\gamma^{a}=2\eta^{ab}$, as follows:
\begin{equation}
\begin{array}{l}
\gamma^{\prime\mu}(x^{\prime})=h^{\mu}_{a}(x^{\prime})\gamma^{a},  \\
\Gamma^{\prime}_{\lambda}(x^{\prime})=\frac{1}{8}[\gamma^{a},\gamma^{b}]g^{\prime}_{\mu\nu}(x^{\prime})h^{\mu}_{a}(x^{\prime}){\cal D}^{\prime}_{\lambda}h^{\nu}_{b}(x^{\prime})
\end{array}
\end{equation}
with ${\cal D}^{\prime}_{\lambda}$ denoting covariant derivatives. Here, we have also $g^{\prime \mu \nu}(x^{\prime})=h^{\mu}_{a}(x^{\prime})h^{\nu}_{b}(x^{\prime})\eta^{ab}$ and $\gamma^{\prime \mu}\gamma^{\prime\nu}+\gamma^{\prime \nu}\gamma^{\prime \mu}=2g^{\prime\mu\nu}(x^{\prime})$. 

It is to be noted that (6) is invariant, by construction, not only under (2), but also under local Galilei or local $G_{5}$ transformations  $\chi^{\prime}(x^{\prime}) \to  T(x^{\prime})\chi^{\prime}(x^{\prime})$, where $ T(x^{\prime})$ corresponds to $x^{\prime \mu}$-dependent $\Lambda^{a}~_{b}$ with $\mu \ne 5$. Incidentally, the latter is a kind of non-Abelian gauge transformations, and $\Gamma^{\prime}_{\mu}$ a gauge-dependent quantity. 

\section{Inertial Forces}
\subsection{Derivation}
First, for the case of $\phi^{\prime}(x^{\prime})$ we may put
\begin{equation}
\phi^{\prime}(x^{\prime})=\exp(-ims^{\prime}/\hbar)\psi^{\prime}(\vec{x}^{\prime},t^{\prime})
\end{equation}
in accordance with (5$^{\prime}$). Substituting this into (5) and using (3), we then find for $\psi^{\prime}(\vec{x}^{\prime},t^{\prime})$:
\begin{equation}
i\hbar\frac{\partial}{\partial t^{\prime}}\psi^{\prime}(\vec{x}^{\prime},t^{\prime})=\Bigl(-\frac{\hbar^{2}}{2m}\vec{\nabla}^{\prime 2}+{\cal H}^{\prime}_{inert}\Bigr)\psi^{\prime}(\vec{x}^{\prime},t^{\prime}) 
\end{equation}
with
\[\hspace{3cm} 
{\cal H}^{\prime}_{inert} \equiv -mR_{jk}\ddot{\tilde{A}}~^{k}x^{\prime j}-i\hbar\dot{R}^{\ell}~_{j}R_{k}~^{j}x^{\prime k}\partial_{\ell}^{\prime} .\hspace{3.5cm} (9^{\prime})
\]
Here we confirm that for $\ddot{\vec{A}}=0$  and $\dot{R}=0$ the equation (9) reduces to the  free-field equation, and ${\cal H}^{\prime}_{inert}$ thus provides a  general expression for inertial forces.

Next, for the case of $\chi^{\prime}(x^{\prime})$ we may write, in terms of 2-component spinors $\psi^{\prime}_{1}$ and $\psi^{\prime}_{2}$,
\begin{equation}
\chi^{\prime}(x^{\prime})=\exp\Bigl(-\frac{im s^{\prime}}{\hbar}\Bigr)
\left(
\begin{array}{l}
\psi^{\prime}_{1}(\vec{x}^{\prime},t^{\prime})  \\
\psi^{\prime}_{2}(\vec{x}^{\prime},t^{\prime}) 
\end{array}
\right),
\end{equation}
and substitute this into (6). By choosing a suitable representation for $\gamma$-matrices, we then find that only $\psi^{\prime}_{1}$ is independent. Now, in proceeding further we have to determine $h^{\mu}_{a}$ (or the gauge). For   $g^{\prime \mu \nu}(x^{\prime})$ given by (3) let us consider the following two cases.

a) We take $h^{\mu}_{a}(x^{\prime })=\partial x^{\prime\mu}/\partial x^{a}$, which are obtained from (2) as
\begin{equation}
\begin{array}{ll}
h^{i}_{j}(x^{\prime})=R^{i}~_{j},& h^{i}_{4}(x^{\prime})=\frac{1}{u}\dot{R}^{i}~_{j}R_{k}~^{j}x^{\prime k}+\frac{1}{u}R^{i}~_{j}\dot{\tilde{A}}^{j},   \\
h^{5}_{i}(x^{\prime})=\frac{1}{u}\dot{\tilde{A}}_{i}, & h^{5}_{4}(x^{\prime})=\frac{1}{u^{2}}R_{j}~^{i}\ddot{\tilde{A}}_{i}x^{\prime j}+\frac{1}{2u^{2}}\dot{\tilde{A}}_{i}\dot{\tilde{A}}^{i},  \\
h^{4}_{4}(x^{\prime})=h^{5}_{5}(x^{\prime})=1 , & h^{i}_{5}(x^{\prime})=h^{4}_{i}(x^{\prime})= h^{4}_{5}(x^{\prime})=0.
\end{array}
\end{equation}
From (4) and (11) it follows, however, that ${\cal D}^{\prime}_{\lambda} h^{\nu}_{b}(x^{\prime})=0$, hence $\Gamma^{\prime}_{\mu}(x^{\prime})=0$. Equation (6) together with (7), (11) and (10) then leads us to
\begin{equation}
i\hbar\frac{\partial}{\partial t^{\prime}}\psi_{1}^{\prime}(\vec{x}^{\prime},t^{\prime})=\bigl(-\frac{\hbar^{2}}{2m}\vec{\nabla}^{\prime 2}+{\cal H}_{inert}^{\prime}\bigr)\psi_{1}^{\prime}(\vec{x}^{\prime},t^{\prime})
\end{equation}
with ${\cal H}^{\prime}_{inert}$ being the same as (9$^{\prime}$).

b) Taking a $t^{\prime}$-dependent $3\times 3$ orthogonal matrix $\bar{R}(t^{\prime})$ we now adopt f$\ddot{{\rm u}}$nfbein $h^{\prime \mu}_{a}$ such as
\[\hspace{3cm} 
\begin{array}{ll}
h^{\prime i}_{j}=\tilde{R}_{j}~^{k}h^{i}_{k}, & h^{\prime 5}_{i}=\bar{R}_{i}~^{j}h^{5}_{j},  \\
 h^{\prime \mu}_{a}=h^{\mu}_{a}& {\rm for~~others}
\end{array}
\hspace{5.2cm} (11^{\prime})
\] 
with $h^{\mu}_{a}$'s given by (11). In this case $\Gamma^{\prime}_{\mu} \ne 0$, and in fact $\Gamma^{\prime}_{i}=0, \Gamma^{\prime}_{5}=0$ and $\Gamma^{\prime}_{4}=\frac{1}{4u}\dot{\bar{R}}~^{\ell}~_{j}\bar{R}_{k}~^{j} \epsilon_{k\ell m}\tilde{\sigma}_{m}$, where $\bar{\sigma}_{m} \equiv {\rm diag}(\sigma_{m},\sigma_{m})$ with $\sigma_{m}$'s being Pauli matrices. After some calculations we then find, for $\psi^{\prime}_{1}(\vec{x}^{\prime},t^{\prime})$, the same equation as (12) but with a further term in the bracket of the right-hand side:
\begin{equation}
{\cal H}^{\prime}_{spin} \equiv \frac{\hbar}{4}\dot{\bar{R}}~^{\ell}~_{j}\tilde{R}_{k}~^{j}\epsilon_{k\ell m}\sigma_{m},
\end{equation}
which is to be regarded as a kind of inertial force.

As remarked earlier, the situation is basically the same for general Bargmann-Wigner fields. Thus, summarizing the above results we can say that when a special $h^{\mu}_{a}$ (or the gauge) is chosen, the inertial forces are all described by ${\cal H}^{\prime}_{inert}$, irrespective of spins.

\subsection{Examples} 
Let us now rewrite ${\cal H}^{\prime}_{inert}$ and ${\cal H}^{\prime}_{spin}$ for some special cases.

  a) ${\cal H}^{\prime}_{inert}$: If we put $R={\rm I}$ and $\vec{A}(t)=\frac{1}{2}\vec{a}t^{2}$ ($\vec{a}$ :const. vector), then
\begin{equation}
\begin{array}{ll}
{\rm 1st~ term} = -m\vec{a}\cdot\vec{x}^{\prime}, & {\rm 2nd ~term}=0,
\end{array}
\end{equation}
as expected.

  b) ${\cal H}^{\prime}_{inert}$: Since  $RR^{tr}={\rm I}$, $\dot{R}R^{tr}$ is a real $3\times 3$ antisymmetric matrix. Thus, in terms of $J_{k}$ such that $(iJ_{k})^{\ell}_{m}=\epsilon_{k\ell m}~(k,\ell,m=1,2,3)$, we can write $\dot{R}R^{tr}\equiv  i\vec{\Omega}(t^{\prime})\cdot \vec{J}$, to find
\begin{equation}
{\rm 2nd~ term}=-\vec{\Omega}(t^{\prime})\cdot \vec{L}^{\prime},~~~~~~\vec{L}^{\prime} \equiv -i\hbar \vec{x}^{\prime}\times \vec{\nabla}^{\prime}.
\end{equation}
For $R(t^{\prime})= \exp (i\vec{\omega}\cdot \vec{J}~t^{\prime}) $ ($\vec{\omega}$ : const. vector), in particular, we have $\vec{\Omega}(t^{\prime})=\vec{\omega}$.

 c) ${\cal H}^{\prime}_{spin}$: Similarly, if we put $\dot{\bar{R}}\bar{R}^{tr}\equiv i\vec{\Omega}^{\prime}(t^{\prime})\cdot \vec{J}$ in ${\cal H}^{\prime}_{spin}$, we find 
\begin{equation}
{\cal H}^{\prime}_{spin}=-\vec{\Omega}^{\prime}(t^{\prime})\cdot \vec{S}, ~~~~~\vec{S}\equiv \frac{\hbar}{2}\vec{\sigma}.
\end{equation}
Thus, if we choose $\vec{\Omega}=\vec{\Omega}^{\prime}$, then the 2nd term in ${\cal H}^{\prime}_{inert}$ plus ${\cal H}^{\prime}_{spin}$ provides $
-\vec{\Omega}\cdot (\vec{L}+\vec{S})$, a result being true only in a special gauge, however.

It is to be remarked here that the above results agree with the Newton approximations to the corresponding cases of general relativity.

\section{Equivalence Principle}
Let us start by considering an inertial system S$_{0}$, where all quantities concerned are denoted by unprimed symbols. The field $\phi(x)$ or $\chi(x)$ then satisfies the free-field equation (5) or (6) with $\Gamma_{\mu}=0$. As shown in [3], the interaction with an external gravitational or Newton potential $\Phi(\vec{x})$ can be introduced there by making a series of replacements: 1) $\eta^{\mu\nu} \to h^{\mu}_{a}h^{\nu}_{b}\eta^{ab}, \gamma^{\mu} \to h^{\mu}_{a}\gamma^{a}$; 2)
\begin{equation}
h^{5}_{4} \to h^{5}_{4}-\frac{1}{u^{2}}\Phi(\vec{x}) ;
\end{equation} 
and then 3) $h^{\mu}_{a} \to \delta^{\mu}_{a}$. Here, the subsidiary condition is left unchanged. The replacements then give
\begin{equation}
\begin{array}{l}
\bigl(\eta^{\mu\nu}\partial_{\mu}\partial_{\nu}+\frac{2}{u^{2}}\Phi(\vec{x})\partial_{5}\partial_{5}\bigr)\phi(x)=0, \\
\bigl(\gamma^{\mu}\partial_{\mu}-\frac{1}{u^{2}}\Phi(\vec{x})\gamma^{4}\partial_{5}\bigr)\chi(x)=0.
\end{array}
\end{equation}
As easily checked, for $\Psi(\vec{x},t)\equiv \psi(\vec{x},t)$ or $\psi_{1}(\vec{x},t)$, each being defined by (8) or (10) unprimed, (18) leads us to
\begin{equation}
i\hbar \frac{\partial}{\partial t}\Psi(\vec{x},t)=\bigl(-\frac{\hbar^{2}}{2m}\vec{\nabla}^{2}+m\Phi(\vec{x})\bigr)\Psi(\vec{x},t).
\end{equation}

We next go over to a non-inertial system S via the transformation (2) with $R(t)={\rm I}$ and $\vec{A}(t)=\frac{1}{2}\vec{a} t^{2}$ ($\vec{a}$: const. vector). That is, we make the replacements: $\partial_{\mu} \to \partial_{\mu}^{\prime}, \eta^{\mu\nu} \to g^{\prime \mu\nu}(x^{\prime})$ and $\gamma^{\mu} \to \gamma^{\prime\mu}$. Note here that $\gamma^{4}=\gamma^{\prime 4}$, $\partial_{5}=\partial_{5}^{\prime}$, and $\Phi(\vec{x})$, being a given external potential , has only to be rewritten in S as $\Phi(\vec{x}^{\prime}-\frac{1}{2}\vec{a}t^{\prime 2})$. Now, for definiteness let us consider the case of uniform gravitation: $\Phi(\vec{x})=\vec{g}\cdot \vec{x}$ ($\vec{g}$: const. vector). Then, the equation for $\Psi^{\prime}(\vec{x}^{\prime},t^{\prime}) \equiv \psi^{\prime}(\vec{x}^{\prime},t^{\prime})$ or $\psi_{1}^{\prime}(\vec{x}^{\prime},t^{\prime})$ turns out, after a phase transformation $\exp(m\vec{a}\cdot \vec{g}t^{3}/6i\hbar)$, to be
\begin{equation}
i\hbar \frac{\partial}{\partial t^{\prime}}\Psi^{\prime}(\vec{x}^{\prime},t^{\prime})=\Bigl[-\frac{\hbar^{2}}{2m}\vec{\nabla}^{\prime 2}+m(\vec{g}-\vec{a})\cdot\vec{x}\Bigr] \Psi^{\prime}(\vec{x}^{\prime},t^{\prime}).
\end{equation}

The above result has an important physical meaning. That is, if we choose a non-inertial system S such as $\vec{a}=\vec{g}$, then the gravitational effect will completely disappear there, being in accordance with Einstein's equivalence principle. The situation is basically the same for general cases of $\Phi(\vec{x})$ and also of Bargmann-Wigner fields.
\section{Conclusions}
 
\hspace{0.5cm}a) Fully utilizing the generally covariant formalism we have derived the general expressions for inertial forces in a purely quantum-mechanical manner.

b) NRQM is compatible with the equivalence principle.

c) In our formalism other related problems can also be discussed within the framework of NRQM, i.e., without recourse to any other theories like classical mechanics.

d) So far as the covariance under space-time transformations is concerned, no difference seems to exist formally between NRQM and RQM.  We may therefore say that the former is a self-supporting, independent theory, which is comparable to the latter.

\begin{reference} 
                  
\item M. Omote, S. Kamefuchi, .Y. Takahashi, Y. Ohnuki,  {\it Fortschr. Phys.,} 37 (1989) 933.
\item D. E. Soper, Classical Field Theory  (John Wiley and Sons, New York, 1976). 
\item M. Omote, S. Kamefuchi,  {\it Proceedings of the International Symposium "Quantum Theory and Symmetries",} 18-22 July 1999, Goslar, Germany, eds. H.D. Doebner et al. (World Scientific, Singapore), to be published.
\end{reference}

\end{document}